\documentclass{mn2e}
\usepackage{epsfig,natbib2,natbibmnfix}
\usepackage{amsmath}
\usepackage{mathrsfs}
\usepackage{subfigure}
\usepackage{url}
\usepackage[dvips]{color}
\usepackage{multicol}

\newcommand{\be}{\begin{equation}}
\newcommand{\ee}{\end{equation}}

\def\ltsima{$\; \buildrel < \over \sim \;$}
\def\simlt{\lower.5ex\hbox{\ltsima}}
\def\gtsima{$\; \buildrel > \over \sim \;$}
\def\simgt{\lower.5ex\hbox{\gtsima}}
\newcommand\sgra{Sgr~A$^*$}

\newcommand\Mdot{\dot{M}}
\def\del#1{{}}

\def\msun{{\,{\rm M}_\odot}}

\title[Modelling supermassive black hole growth]{Modelling supermassive black hole growth: towards an improved sub-grid prescription}

\author[Alexander Hobbs, Chris Power, Sergei Nayakshin \& Andrew R. King]
       {\parbox{18cm}{Alexander Hobbs$^{1}$, Chris Power$^{2}$, Sergei Nayakshin$^{3}$ \&
           Andrew R. King$^{3}$}\vspace{0.3cm}\\
         \noindent $^{1}$Institute for Astronomy, ETH Zurich $^{2}$University of Western Australia $^{3}$Dept. of Physics \& Astronomy, University of Leicester}

\begin{document}

\maketitle

\begin{abstract}
Accretion onto supermassive black holes (SMBHs) in galaxy formation simulations is frequently modelled by the Bondi-Hoyle formalism. Here we
examine the validity of this approach analytically and numerically. We
argue that the character of the flow where one evaluates the gas properties is unlikely to satisfy the simple Bondi-Hoyle model.
Only in the specific case of hot virialised gas with zero angular momentum and negligible radiative cooling is the Bondi-Hoyle solution relevant.
In the opposite extreme, where the gas is in a state of free-fall at the evaluation radius due to efficient cooling and the dominant gravity of
the surrounding halo, the Bondi-Hoyle formalism can be erroneous by orders of magnitude in either direction. This may impose artificial
trends with halo mass in cosmological simulations by being wrong by different factors for different halo masses. We propose an expression for the sub-grid accretion rate
which interpolates between the free-fall regime and the Bondi-Hoyle regime, therefore taking account of the contribution of the halo to the gas dynamics.
\end{abstract}

\begin{keywords}{}
\end{keywords}
\renewcommand{\thefootnote}{\fnsymbol{footnote}}
\footnotetext[1]{E-mail: {\tt ahobbs@phys.ethz.ch}}

\section{Introduction}

Over the last decade, compelling observational evidence has revealed that
many galaxies in the local Universe harbour supermassive black holes (SMBHs) 
with masses $10^6 \simlt M_{\rm bh}/{\rm M}_{\odot} \simlt 10^9$ in their 
centres. 
During the same period, surveys of the distant Universe uncovered the
existence of quasars
at $z\sim 7$, when the Universe $\simlt 1/10^{\rm th}$ of its current age;
this implies that many SMBHs had already assembled 
their mass by this time \citep{MortlockEtal2011}.

Our understanding of the physics that dictates the growth of SMBHs is
incomplete. Black holes grow by accreting 
low angular momentum material from their surroundings, yet the character
of the accretion flow onto an SMBH is governed by physical
processes as diverse as galaxy mergers \citep[e.g.,][]{HopkinsQuataert2010}, 
turbulence induced by stellar feedback \citep[e.g.,][]{HobbsEtal2011} and 
black hole accretion-driven outflows \citep[e.g.,][]{nayakshin.power.2010}.

Black hole growth is now routinely modelled in galaxy formation simulations
\citep[for a fiducial work see][]{SpringelEtal05} and the importance of SMBHs in shaping the
properties of galaxies is now well established
\citep[e.g.][]{croton.etal.2006,bower.etal.2006}.  The majority of galaxy
formation simulations published in the literature incorporate what we shall
term the ``Bondi-Hoyle model" for black hole growth
\citep[see, e.g.,][]{SpringelEtal05, SijackiEtal2007, PelupessyEtal2007, DiMatteoEtal2008, JohanssonEtal2009, KimEtal2011}, which derives from the work of \citet{Bondi44}
and \citet{Bondi52} - hereafter B\&H. This model assumes the simplest possible accretion flow,
where the gas is at rest at infinity and accretes steadily onto a black hole,
subject only to the (Newtonian) gravity of the latter, which is modelled as a
point mass.

Simulations that model the idealised physical problem as it is set out in
B\&H, or in idealised generalisations, produce results that are in good agreement
with the analytical solution \citep{Ruffert94, BaraiEtal2011}. In galaxy formation
simulations, unfortunately, this idealised picture is far from satisfied, 
as the gas inflow is complicated considerably by the properties of the flow at larger scales. 
The most notable example of this is the presence of non-zero angular momentum 
that provides a natural barrier to eventual accretion by the SMBH.
Gas settles into a disc whose dimensions are set by the angular momentum of the accretion flow,
with only the very lowest angular momentum material able to accrete. A true estimate of the
accretion rate onto the SMBH must therefore take account of this angular momentum, and indeed attempts to include it in an accretion sub-grid model have been made \citep{PowerEtal2010, LevineEtal2010, DeBuhrEtal2011a} along with the presence of turbulence and/or vorticity in the gas \citep{KrumholzEtal05, KrumholzEtal06} finding large departures from the standard Bondi-Hoyle rate.

In this short paper we wish to make a simple and more fundamental point
that in galaxy formation simulations even spherically symmetric accretion cannot be correctly modelled by the Bondi-Hoyle formalism,
except in the most specific of cases. To make this point we suspend, for the moment, our
disbelief that gaseous infall can proceed entirely radially from large scales and consider zero angular momentum
accretion flows onto an SMBH embedded in the potential of a massive dark matter halo. Indeed, this
is an example of a situation where one might expect the B\&H formula to provide a reasonable
estimate of the accretion rate.

The layout of this paper is as follows. In Section 2 we show analytically that 
in large-scale simulations of cosmological volumes the Bondi-Hoyle approach is 
invalid, and in Section 3 we present some numerical tests of this hypothesis. 
Finally in Sections 4 \& 5 we discuss our conclusions.

\section{Analytical arguments}

\subsection{Classical Bondi-Hoyle accretion}
\label{sec:bondistart}

We first recap the main assumptions underpinning the classical B\&H 
papers. These are nicely summarised in the first sentence of the 
\cite{Bondi52}'s abstract: ``The special accretion problem is investigated 
in which the motion is steady and spherically symmetrical, the gas 
{\em being at rest at infinity}''. We have italicised the part of the 
sentence that bears the most importance for us here. 

Physically, gas can be at rest at infinity only when it is not subject to any
forces. The only external force acting on the gas in the restricted 
B\&H problem, i.e., the gravitational force, is due to
the black hole. Self-gravity of the gas is neglected. The ``infinity'' in question is a
region at a distance large enough from the SMBH that the gravitational force exerted by the latter
is negligible when compared to the pressure forces within the gas. This is quantified by
defining the Bondi (or the ``capture'') radius,
\begin{equation}
r_B = 2GM_{\rm BH}/c_\infty^2
\end{equation}
where $M_{\rm BH}$ is the mass of the central object and $c_\infty$ is the
sound speed of the gas far from the hole. The Bondi radius divides the flow into two 
distinct regions \citep{Frank02}. Far from $r_B$, gas is hardly aware of the 
existence of the black hole, and the flow is very subsonic. The pressure and 
density of a subsonic flow are approximately constant, therefore we can set 
$\rho(r) \approx \rho_\infty$ at $r\gg r_B$.

Inside the capture radius, on the other hand, $\rho(r)$ begins to increase 
above the initial value, and the flow eventually reaches a sonic point where 
$|v_r| = c_\infty$, within which it plunges essentially at free-fall. The
sonic point is found from $r_{\rm s} = GM_{\rm BH}/2c^2_{\rm s}(r_{\rm s})$, where 
$c_{\rm s}(r_{\rm s})$ is the sound speed at $r_{\rm s}$. This local quantity\footnote{note that $r_{\rm s}$ here is the sonic radius, not the scale radius as it is commonly used in descriptions of dark matter halo profiles} 
is related to the sound speed at infinity \citep{Frank02} via $c_{\rm s}(r_{\rm s}) = c_\infty (2/(5-3\Gamma))^{1/2}$
where $\Gamma$ is the polytropic index of the gas that relates the gas 
pressure and density by $P = K \rho^\Gamma$, with $K$ a positive constant. 

Applying the Bondi-Hoyle formalism to black hole growth assumes that the 
accretion rate onto the SMBH is commensurate with the accretion rate through 
the Bondi radius (i.e., the flow is steady-state) and therefore given by  
\begin{equation}
\dot M_{\rm BH} = \pi \lambda(\Gamma) r_B^2 \rho_\infty c_\infty = \frac{4 \pi
  \lambda(\Gamma) G^2 M_{\rm BH}^2 \rho_\infty}{c_\infty^3}
\label{eq:mdt}
\end{equation}
where $\lambda(\Gamma)$ contains all the corrections arising due to the finite
pressure gradient force in the problem. This function varies relatively
weakly, i.e., between $1.12$ for $\Gamma=1$ and $0.25$ for $\Gamma = 5/3$ \citep{Bondi52}. For 
the remainder of this paper our fiducial assumption is a soft equation of
state i.e., $\Gamma \approx 1$.

\subsection{When is Bondi-Hoyle accretion applicable?}
\label{sec:bondianalytical}

The Bondi-Hoyle formalism has been widely adopted as a `sub-grid'
prescription for the accretion rate onto the SMBH in large-scale cosmological
simulations. The standard argument is that while one cannot usually resolve
the scales of the Bondi radius, one can at least use the smallest resolved
scales to approximately determine the value of the gas density and the sound
speed ``at infinity'' for use in the Bondi formula \citep[see, e.g.,][]{BoothSchaye2009}. Indeed, the smallest
resolvable scales are usually about a fraction of a kpc, whereas the Bondi
radius is of the order of a few to a few tens of pc.

For the \cite{Bondi52} solution to be applicable even to spherical flow, we need to
make sure that the ``gas \emph{being at rest at infinity}'' assumption is
satisfied where the relevant gas properties (density and sound speed) are evaluated. In cosmological simulations we expect SMBHs to be immersed in stellar
bulges and dark matter haloes that are typically $\sim 10^3$ to $10^4$ times
more massive than the SMBH \citep[see, e.g.,][]{Haering04, GuoEtal2010}. If gas in the halo (or bulge) is as hot as the
halo virial temperature, it will be in hydrostatic balance \citep[see, e.g.,][]{KomatsuSeljak2001, SutoEtal98}. It seems that in this
situation, which is common for low luminosity SMBHs in giant elliptical
galaxies where the gas is rather tenuous and hot since the cooling time is long \citep{ChurazovEtal2005} the Bondi-Hoyle solution is potentially useful.

However, in the epoch when SMBHs grow rapidly, their hosts are very gas
rich, and the inflow of gas from large scales cannot be easily captured by gradual cooling from a tenuous hot halo \citep[see, e.g.,][]{BirnboimDekel2003, KeresEtal2005, KeresEtal2009, DekelEtal2009, KimmEtal2011}. Higher density gas is likely to cool much faster and hence is
likely to be much cooler than the virial temperature. In this case the gas is
not able to support its own weight, and must collapse to the centre, where it
feeds the SMBH and forms stars. Therefore we expect a radial inflow of gas to
the centre rather than a hydrostatic balance ``far'' from the
SMBH. We note that \cite{Ricotti07} have demonstrated how the Bondi-Hoyle formalism must be modified
in the presence of an external potential for the specific case of the growth
of primordial black holes in a dark matter halo with a power-law density
profile. 

In actuality, even the meaning of the Bondi radius becomes unclear in this
situation, as gas is not virialised near the SMBH, and the halo potential plays an important role. In the ``naked SMBH'' problem, $r_B$
delineates the region inside which the potential energy of the hole starts to become
greater than the internal energy of the gas. For an SMBH plus host halo
system, one should introduce a modified Bondi radius,
\begin{equation}
\tilde{r}_{B} = 2GM_h/c_\infty^2
\end{equation}
that takes into account the total mass of the halo. In order for the gas to
accrete efficiently onto a dark matter halo, its temperature must be at most
comparable with the virial temperature at the outer edge of the halo
\citep{WhiteFrenk1991}. Thus, we set $c_\infty \simlt GM_h / r_h$, where $r_h$ is the halo virial radius, giving us
a modified Bondi radius of $\tilde{r}_B \simgt 2r_h$. The gravitational
potential energy starts to dominate the internal energy of the infalling gas
before the latter has even reached the edge of the halo, and so the standard $r_B$ is meaningless in this case.

\begin{figure}
\begin{minipage}[b]{.49\textwidth}
\centerline{\psfig{file=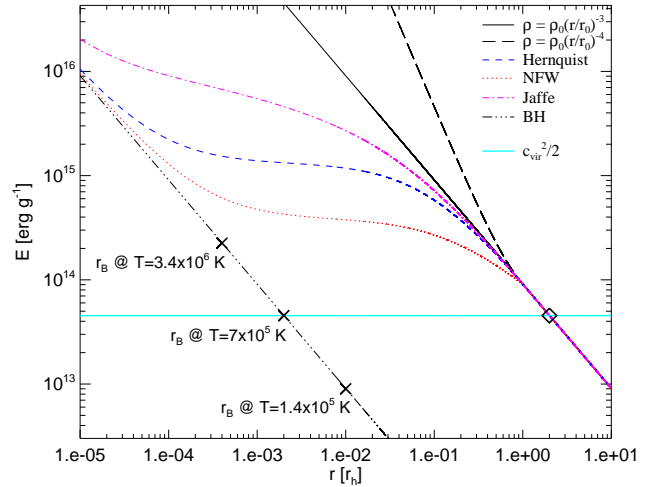,width=1.00\textwidth,angle=0}}
\end{minipage}
\caption{Plot of potential energies for a selection of dark matter halo
  profiles (some realistic, some instructive) modified by a central SMBH,
  compared to the thermal energy of the gas at the virial temperature at the
  halo virial radius. The halo profiles are normalised to contain a mass $M_h = 10^{11} \msun$ within the halo radius $r_h$, outside of which the potential is Keplerian by construction. The potential due to the central SMBH by itself is also
  plotted. The classical Bondi radius for three different values of
  temperature are indicated as crosses - the greater the separation
between the cross denoting $r_{B}$ and the curves of potential energy for a given halo
the more inaccurate the standard Bondi-Hoyle approach. It is clear then that the latter becomes more accurate at higher $T$.
The modified Bondi radius for the
  fiducial $T_{\rm vir}$ is denoted by a rhombus, demonstrating that at
  this temperature the potential energies are dominant for the entirety of the halo.}
\label{fig:phi_all}
\end{figure}

Figure \ref{fig:phi_all} illustrates this point graphically by comparing the
potential energy of gas as a function of radius for a variety of halo profiles
with $c_{\rm vir}^2/2$, assuming that the gas temperature is virial at $r_h$. These profiles are
described in the Appendix. The cosmology we have assumed for the halos is
$\lambda \rm{CDM}$ with $\Omega_{\rm m} = 0.27$, $\Omega_{\lambda} = 0.73$ and
a virial overdensity parameter of $\Delta = 200$ at a redshift of $z = 2$
(although we note that our conclusions are unchanged for a wide range in $z$,
from the early Universe to the present day).

The $\propto 1/r$ potential due to the SMBH of mass $M_{\rm BH} = 10^8 \msun$
is shown with the dash-triple-dot power law. The traditional Bondi (or capture) radius is at $r = 0.002$, where the SMBH potential crosses the
$c_{\rm vir}^2/2$ line. For all the halos considered, the potential energy of the host
makes a significant contribution at all radii except those within the very
inner parts of the halo, exceeding the gas internal energy by a large factor everywhere outside the SMBH
radius of influence.

Therefore, if we assume that the gas has a relatively soft equation of state and
accretes spherically onto the halo at or below $T_{\rm virial}(r_h)$, the character of the inflow becomes that of supersonic
inside the halo \emph{instead of being stationary at infinity}.  In the
Appendix we quantify this by calculating the sonic point for isothermal gas
flows at or below the virial temperature and for all of the halo mass profiles
considered. We show that for typical dark matter halos the sonic point is
reached while still at very large distances from the central black hole.

From Figure \ref{fig:phi_all} it can also be seen that for hotter gas, the
classical $r_B$ estimate becomes more accurate, as the
potential energy of the (halo + SMBH) system starts to asymptote to the SMBH
solution  and no longer dominates over the
value of $c_{\rm vir}^2/2$. If the gas thermal energy is comparable to that of the
potential energy we naturally expect near hydrostatic equilibrium to be
maintained and in this case the Bondi-Hoyle formalism is applicable to spherical flow.

Of course, this is a simplified picture. In reality gas that has accreted onto a halo from outside the virial radius may shock at smaller radii, heating up to the local $T_{\rm vir}$ at that radius. In the `cold mode' where the galaxy is assembled via cold streams that penetrate far inside the halo, such shock heating is likely to occur when the infalling gas reaches the radius of the galactic disc, at a small fraction of $r_h$ \citep{KeresEtal2005}. However, due to the high densities reached for the shocked gas the cooling time is likely to be short \citep{BirnboimDekel2003, KeresEtal2005}, and in particular in the presence of Compton cooling from a quasar radiation field it will be significantly less than the free-fall time at that radius \citep{NulsenFabian2000}. Gas may therefore be stationary temporarily at the shock radius but as it cools and begins to infall it will quickly tend to the free-fall velocity, and certainly by the time a radial inflow reaches the classical Bondi radius - values for which can be seen in Figure \ref{fig:phi_all} for different temperatures - the assumption of being at rest will no longer be satisfied.


Put simply, this is an energy argument. Regardless of where the gas might begin to infall from, the halo imposes a far stronger gravitational potential energy than the SMBH everywhere outside of the SMBH radius of influence (by definition). The potential energy due to a typical halo also increases to smaller radii from any point at which the gas is likely to have reached the virial temperature (see Figure \ref{fig:phi_all}). As a result it is clear that gravity will often dominate over thermal energy at scales significantly larger than the classical Bondi radius for the SMBH.

In the interest of completeness we now demonstrate numerically the form of the radial infall in a realistic background potential within a galaxy. We choose a dynamic range that
lies inside the sonic point (as would always be the case for efficiently
cooled gas at these scales, as we have shown in the Appendix) in order to highlight why the standard Bondi estimate is inaccurate here. 

\section{Numerical tests}\label{sec:numerics}

To perform the simulations we employ the three-dimensional smoothed particle hydrodynamics (SPH)/N-body code
GADGET-3, an updated version of the code presented in
\cite{Springel05}. The gas is evolved in a static external potential that includes a point mass black hole at
the centre. The computational domain extends from a kiloparsec down to an ``accretion radius''
around the black hole at $r_{\rm acc} = 1$ pc, and we remove the particles
that come within this distance of the SMBH. 

For the external potential in our model we use a Jaffe cusp as per equation \ref{eq:gammas} but with a core at the centre of our computational domain to prevent divergence in the gravitational force. The radius of the core, $r_{\rm c}$, corresponds to
approximately the dynamical influence radius of the SMBH. With this (modified) potential
the mass enclosed within radius $r$ is given by:
\begin{equation}
M(r) \; = \; M_{\rm bh} + \;
\begin{cases}
M_{\rm c} \left(\frac{r}{r_{\rm c}}\right)^3 \;, &r < r_{\rm c}\\
M_{\rm c} + a M \left(\frac{1}{r_{\rm c} + a} - \frac{1}{r + a}\right), &r \ge r_{\rm c}\;,
\end{cases}
\label{eq:potential}
\end{equation}
where $M_{\rm c} = 2 \times 10^8 \msun$, $M = 10^{11} \msun$,
$r_{\rm c} = 20$ pc, and $a = 10$ kpc. The mass of the
SMBH is set to $M_{\rm bh} = 10^{8} \msun$.

For simplicity, the gas is kept isothermal throughout the entirety of the simulation and self-gravity is turned off.

\subsection{Initial conditions}\label{sec:ic}

The starting condition for our simulations is that of a uniform density,
spherically symmetric thick gaseous shell with mass $M_{\rm shell} = 10^8 \msun$, that ranges from $r_{\rm in} = 0.1$ kpc to $r_{\rm out} =
1$ kpc and is centered on the black hole. The temperature of the gas is varied between tests, 
ranging from $10^3$ K to $10^5$ K. To minimise initial
inhomogeneities we cut the shell from a relaxed, glass-like configuration. 

The gas infalls from rest within our static
potential.  After a time of the order of the dynamical time at the outer edge
of the shell, a steady-state radial mass flux is reached for the majority of the gas and it is at this time that we
make our comparisons between the various accretion rates in the next section.

\subsection{Results}\label{sec:results}

We define a radius-dependent ``measured'' accretion rate as $\dot M(r) = 4\pi
r^2 \rho v_r$, where $v_r$ is the radial velocity of the gas.
Since the system has been allowed to settle into an approximate steady-state,
this function (shown as an average by the long dashed red line) is almost constant with
radius, and is the same as the time-averaged accretion rate measured at the
black hole.

The solid, the dotted and the dashed curves in Figure \ref{fig:bondiff} show
the standard Bondi-Hoyle estimate for the accretion rate (equation \ref{eq:mdt}) as a function of radius for
three different values of gas temperature. Each of these curves uses a `local' $\rho$,
the density at each radius, in order to represent where the $\rho_\infty$ might be evaluated in a large-scale simulation.

Clearly, as Figure \ref{fig:bondiff} shows, the Bondi-Hoyle estimate is very
inaccurate for these isothermal simulations. At intermediate temperatures,
e.g., $10^5$ K, the formula in the inner parts results in a significant
overestimate of the accretion rate and an underestimate at large radii.

\subsubsection{Free-fall rate}

A simple but physically well motivated alternative to the Bondi-Hoyle formula
for efficiently cooled spherically-symmetric flows is a free-fall rate estimate, 
\begin{equation}
\Mdot_{\rm ff}(r) = \frac{M_{\rm gas, enc}(r)}{t_{\rm ff}(r)}\;,
\label{eq:ffmgasenc}
\end{equation}
where $M_{\rm gas,enc} (r)$ is the enclosed gas mass within radius $r$, and $t_{\rm
  ff} = (r^3/2GM(r))^{1/2}$ is the free fall time. Alternatively one may wish to
approximate the enclosed gas mass in the above equation as $(4\pi/3) r^3
\rho_{\rm gas}(r)$, so that
\begin{equation}
\Mdot_{\rm ff}(r) \sim \frac{4 \pi r^3 \rho_{\rm gas}(r)}{3 t_{\rm ff}(r)}
\label{eq:ffrho}
\end{equation}

\begin{figure}
\begin{minipage}[b]{.49\textwidth}
\centerline{\psfig{file=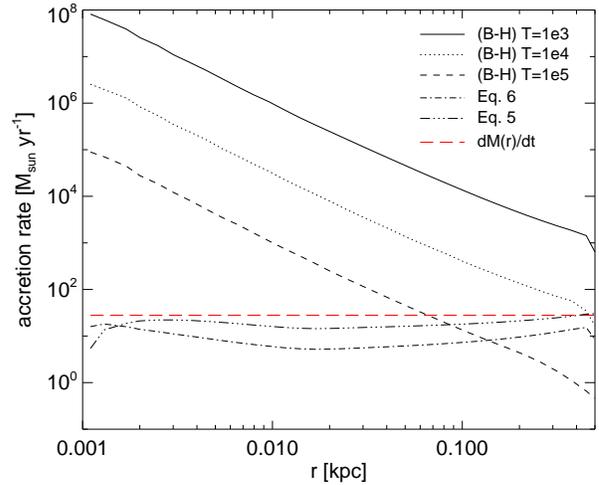,width=1.00\textwidth,angle=0}}
\end{minipage}
\caption{The mass flux (steady-state value) from the simulations through each
  radius (red) i.e., $\Mdot(r)$, together with the Bondi-Hoyle estimate as per
  Equation \ref{eq:mdt} and the free-fall estimate as per Equations
  \ref{eq:ffmgasenc} \& \ref{eq:ffrho}.}
\label{fig:bondiff}
\end{figure}

Both of the above estimates provide a much better match to the accretion rate
in our numerical tests than the Bondi-Hoyle approach, as Figure
\ref{fig:bondiff} shows. It should be noted too that since the
free-fall estimates have no dependance on $c_{\rm s}$, the profiles are
converged regardless of the temperature.

\section{Discussion}\label{sec:discussion}

Based on our analytical arguments and numerical tests, we conclude that the \citet{Bondi52} formula, designed for accretion onto ``naked
black holes'', can only be applied to accretion onto astrophysical SMBHs,
i.e., those embedded in massive dark matter halos, if the gas in the
latter is at or near hydrostatic equilibrium.  It is only in this case that one of the key
assumptions of \citet{Bondi52} -- the ``gas being at rest at infinity'' -- is
satisfied (infinity meaning outside the Bondi radius). If the gas cooling time is long then
it is possible that this state may be reached, and indeed this is probably the
case in giant gas-poor elliptical galaxies, where the gas is tenuous and hot \citep[see, e.g.,][]{ChurazovEtal2005}. 

However, in the most interesting phase of SMBH and galaxy buildup, when the
halo is likely to be awash with gas to feed both the SMBH and star formation,
densities are high and the cooling time is expected to be short, with gas cooling far below the
virial temperature. Hydrostatic balance is then extremely unlikely for gas in the
halo, and if it is not supported by either shock heating or angular momentum it will tend to free-fall velocities. We have run a series of simple numerical
tests to explore this limit, allowing a thick, spherical shell of gas to
accrete onto an SMBH at the centre of a background halo potential.  These
tests showed that a free-fall accretion rate estimate is indeed much more
accurate than the Bondi-Hoyle formalism in this case. What is most concerning
is that the error of the latter strongly depends on gas temperature (and thus
the cooling function) and cannot be ``predicted''. The Bondi-Hoyle accretion
formalism may thus be wrong by significant (and unknown) factors in either
direction. The error may well depend on the galaxy mass and type
systematically. Predictions based on the Bondi-Hoyle formalism alone are therefore unlikely to
be robust. 

We note too that this problem persists even for simulations that are able to resolve down to the classical Bondi radius of the SMBH.
While the properties of the gas (density, sound speed) evaluated at this radius may be correct in the sense that they do not suffer from resolution problems,
it is unlikely that this radius will be a true ``infinity'' where the gas is in hydrostatic equilibrium - again, the reason for this is
the influence of the more massive background halo potential on the dynamics of the gas.

For realistic flows of course the situation is more complex. Here we find that the best way to present our argument is in terms of ``supporting'' mechanisms for the gas at the radius where one evaluates the gas properties in a simulation. To start with, we have the two extremes:
(i) Bondi-Hoyle, where the gas is completely supported and stationary at the evaluation radius;
and (ii) free-fall, where the gas is entirely radial and influenced only by gravity i.e., has zero support.
It is important to recognise that the reality will lie somewhere in-between. Specific cases that lie between these extremes are:
(iii) the gas is unsupported at the evaluation radius but is shocked at smaller radii, after which it may begin to tend to free-fall
once again if the cooling time is short, and (iv) the gas possesses angular momentum, and is thus supported by rotation but inflows through viscous processes.

What we therefore desire is a formula for the sub-grid accretion rate that interpolates between the two extremes based on the relevant supporting mechanism.
As we mentioned in the introduction, angular momentum concerns are extremely important in determining the correct accretion rate; and indeed the picture
can be further complicated by the effect of star formation, where it is not clear whether forming stars would deprive the SMBH of fuel \citep[as may have been the case with \sgra - see, e.g.,][]{NC05}, or if feedback
from star formation would actually enhance accretion through a broadening of the angular momentum distribution \citep{HobbsEtal2011}.
Feedback from the AGN itself must also be a consideration, and in particular the interplay between SMBH feeding and AGN feedback, due to the fact that 
these processes connect the small to the large scales. Indeed, a number of authors have conducted detailed investigations into the multi-scale coupling 
between feeding and feedback \citep[see, e.g.,][]{CattaneoTeyssier2007, DuboisEtal2010, KimEtal2011}, often finding that SMBH growth enters into a self-regulated state that oscillates between periods of high and low activity. We plan to investigate all of the processes relevant to SMBH growth and thereby to further develop the sub-grid model; for now, however, we focus on a 
direct interpolation between the Bondi-Hoyle regime and the free-fall regime as a starting point.




The formula we propose is a minor modification of the full Bondi-Hoyle-Lyttleton \citep{HoyleLyttleton1939, Bondi52} expression for an accretor that is moving relative to the gas, namely
\begin{equation}
\dot M_{\rm BHL} = \frac{4 \pi \lambda(\Gamma) G^2 M_{\rm BH}^2 \rho_\infty}{(c_\infty^2 + v_{\rm rel}^2)^{3/2}}
\label{eq:mbhl}
\end{equation}
where $v_{\rm rel}$ is the relative velocity between the SMBH and the gas, which in purely radial flow is zero. In order to interpolate between this and the free-fall rate (equation \ref{eq:ffmgasenc}) we make two changes: (1) replace the relative velocity with the velocity dispersion for the external potential, $\sigma \sim (GM_{\rm enc}(r)/r)^{1/2}$, and (2) replace the black hole mass with the enclosed mass of the external potential, $M_{\rm enc}(r)$. The resulting expression,
\begin{equation}
\dot M_{\rm interp} = \frac{4 \pi \lambda(\Gamma) G^2 M_{\rm enc}^2 \rho_\infty}{(c_\infty^2 + \sigma^2)^{3/2}}
\label{eq:minterp}
\end{equation}   
tends to the standard Bondi-Hoyle formula (equation \ref{eq:mdt}) in the limit that $c_\infty \gg \sigma$, i.e., the gas internal energy dominates over the potential energy of the halo, and $M_{\rm enc} \rightarrow M_{\rm BH}$, i.e., as we approach the SMBH radius of influence. In the opposite limit, where the halo potential energy dominates, $\sigma \gg c_\infty$, we recover the free-fall estimate, equation \ref{eq:ffmgasenc}. Finally, in order to incorporate (in a crude fashion) the effect of AGN feedback on the accretion the rate should be capped at Eddington, namely
\begin{equation}
\dot M_{\rm acc} = \text{max}(\dot M_{\rm interp}, \dot M_{\rm Edd})
\end{equation}
as is commonly employed in the majority of the simulations of SMBH growth that we have mentioned so far.

We hope that with this interpolated expression the sub-grid SMBH accretion rate will take account
of the presence of the halo and indeed any other external potential (e.g. stellar bulge, stellar halo, etc.) that
contributes significantly to the dynamics of the infalling gas. We note that some simulations may already use a velocity dispersion as a cap on the $v_{\rm rel}$ parameter
in order to avoid excessive relative velocities as a result of poor resolution \citep[e.g.,][]{DuboisEtal2010}, in which case the change to the sub-grid expression is minimal.

This interpolation approach is particularly relevant to the picture of the `hot' and `cold' modes of accretion in galaxy formation, that emerged in the 1970s with the paradigm of hot halo gas that is at or near $T_{\rm vir}$ after being shock heated at the virial radius \citep[see the original papers by e.g.,][]{ReesOstriker1977, Silk1977, WhiteRees1978}, and has since been adjusted by results from simulations \citep[although was suggested also in the 1970s based on analytical arguments - see][]{Binney1977} to include cold gas with a soft equation of state that penetrates down to small scales through streams and filaments \citep[see, e.g.,][]{KatzEtal1994, FardalEtal2001, KeresEtal2005, KeresEtal2009}.

There is typically a halo mass scale which delineates the relative importance of each mode, although this varies somewhat in the literature, with \cite{BirnboimDekel2003} finding $M_{\rm halo} \sim 10^{11}$ as the result of 1-D numerical calculations and \cite{KeresEtal2005} finding a factor of 2-3 higher from fully 3-D N-body/hydrodynamical simulations. However, in a subsequent paper, \cite{KeresEtal2009} find that although this transition mass marks the point above which hot, virialised gas atmospheres develop in halos, the actual contribution to the accretion rate to small scales is still dominated by the cold filamentary mode even at higher mass, with hot mode accretion only starting to become important at late times $(z < 1)$. In addition, there is (indirect) observational evidence for cold accretion flows in galaxies through the HI column density distribution \citep{vandeVoortEtal2011}. This suggests that utilising purely a `gas being fully supported' Bondi-Hoyle accretion prescription is unlikely to capture enough of the relevant accretion behaviour in galaxy formation. Using the interpolation expression above would automatically adjust for the hot and cold mode dominance, via the relative importance of the $c_\infty^2$ and $\sigma^2$ terms in the denominator. We note that semi-analytic models of galaxy formation often employ a distinction in the treatment of accretion rate between the two modes \citep[see, e.g.,][]{HirschmannEtal2011}.


We would like, however, to emphasise again the most important caveat with the picture we have presented - the lack of angular momentum. 
If indeed the free-fall mode of accretion becomes more dominant as a result of employing the interpolated accretion rate expression then one must be 
very careful, for it is possible that the residual angular momentum of gas infalling at free-fall velocities from large scales becomes even more relevant to the accretion 
rate \citep{NulsenFabian2000}, due to the formation of a centrifugally supported disc at small scales. In this case we argue that any expression 
which does not take account of angular momentum should be viewed as more of a `capture' rate, with the actual accretion onto the SMBH modelled with a 
viscous timescale, the parameters of which are set by the properties of the gas at the transition point between `infall' and `disc-mode' provided 
it is sufficiently resolved \citep[see][]{PowerEtal2010}. We note for completeness however that an alternative picture where the formation of a disc is not in fact
a hinderance to SMBH growth was presented by \citet{MayerEtal2010}, who found that global instabilities in the disc-like structure that formed from a merger    
led it to collapse and feed the SMBH on a dynamical timescale \citep[and see also the multi-scale simulations by][]{HopkinsQuataert2010}.




\section{Conclusion}

In this paper we have shown that the de-facto industry standard -- the Bondi-Hoyle formalism for accretion rate onto the SMBH fails for more than one reason in a realistic cosmological simulation whenever gas cooling is efficient. We have shown that a free-fall estimate is more appropriate in this case (provided that angular momentum does not impede accretion of gas even further). We suggested an approximate interpolation formula that bridges the rapidly cooling and the inefficient cooling regimes which we hope will be useful for cosmological simulations that cannot resolve gas flow all the way down to the Bondi radius (a few to few tens of parsecs).

Finally, we conclude with the answer to the question
posed in the paper title, which we feel is worth re-iterating: what matters in a simulation is the
\emph{character} of the flow where one evaluates the accretion/capture rate. If it is (a) spherical, (b) at rest,
and (c) sufficiently far away to count as `infinity', with (d) the enclosed mass dominated by the SMBH, and finally
(e) the flow is uninterrupted between the evaluation radius and the black hole, then and only then is
Bondi-Hoyle accretion applicable as a sub-grid model. As we have mentioned, these latter requirements may be met to a sufficient degree by massive gas-poor
ellipticals if sufficiently resolved in a simulation, but are not by any situation where there is appreciable infall from large scales. We note that
the interpolation formula we have proposed is applicable to both cases, although requires further development in order to take account of the full range of accretion regimes as discussed.

\section{Acknowledgments}

We thank the referee for a detailed report that helped improve the paper. We acknowledge useful discussions with Justin Read, and thank Volker Springel for the version of GADGET-3 that was used for the simulations. Theoretical Astrophysics at the University of Leicester is supported by an STFC rolling grant.

\appendix

\section{The sonic point}

\begin{figure}
\begin{minipage}[b]{.49\textwidth}
\centerline{\psfig{file=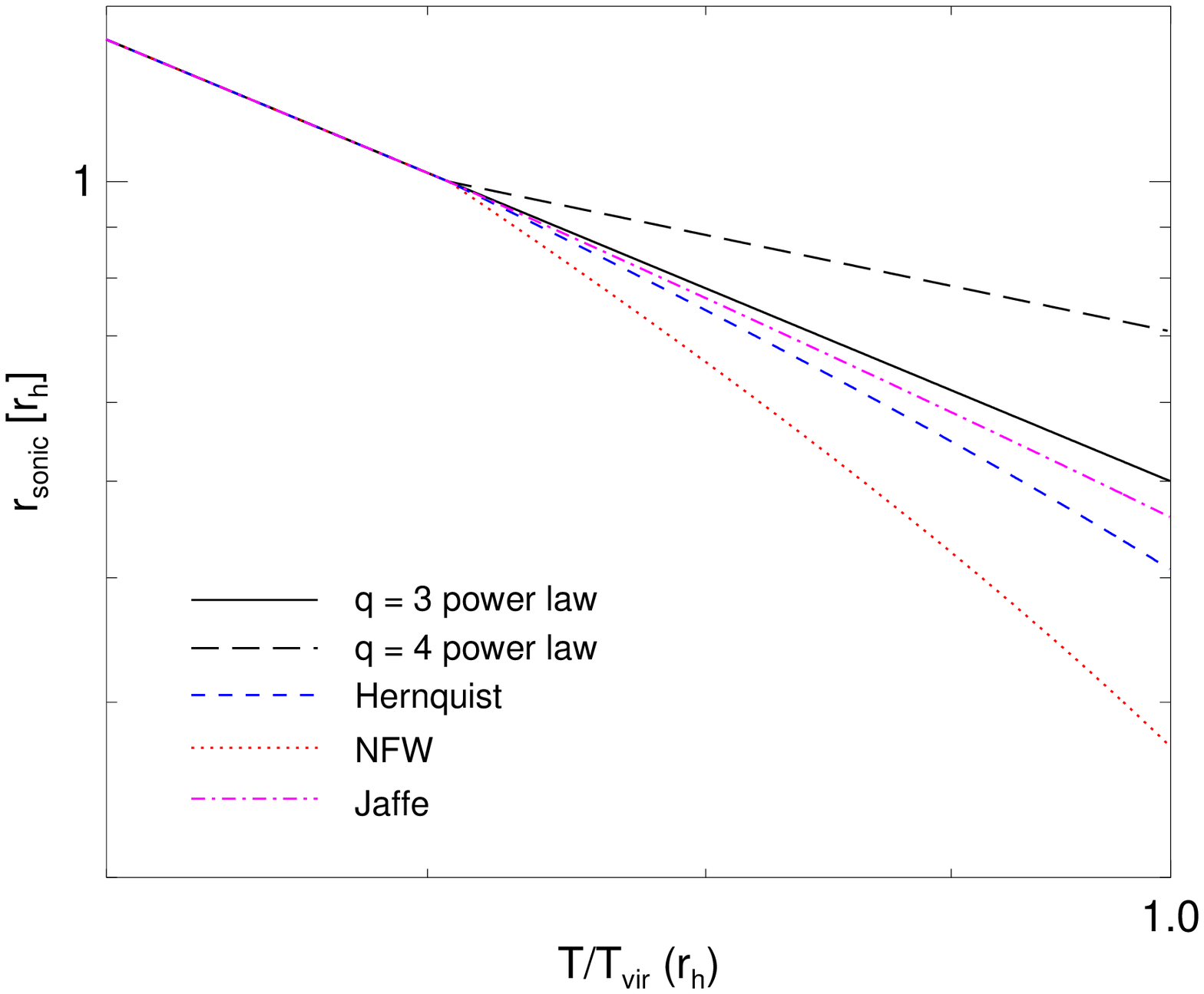,width=1.00\textwidth,angle=0}}
\end{minipage}
\caption{Sonic radius, scaled in units of $r_{\rm h}$, for a variety of halo
  profiles, assuming a halo of $10^{11} \msun$.}
\label{fig:r_sonic}
\end{figure}

For an isothermal gas flow, the sonic point for an extended mass distribution
satisfies $r = GM(r)/2c_s^2(r)$, where $M(r)$ is the enclosed mass. Depending
on the mass profile, this may be inside or outside the halo. We now consider
several profiles for dark matter halos, solve for the sonic point, and plot
its location versus gas temperature in Figure \ref{fig:r_sonic}.  In
particular, for a power-law density profile with index $q$,
\begin{equation}
M(r) = M_{\rm h} \left(\frac{r}{r_{\rm h}}\right)^{3-q}
\label{eq:powerlaw_menc}
\end{equation}
so that $r_{\rm s} = 2^{1/(2-q)} r_{\rm h}$ for gas at $T_{\rm vir}(r_{\rm
  h})$. The enclosed mass in the Navarro, Frenk and White (NFW) profile
\citep{NFWpaper} is
\begin{equation}
M(r) = 4 \pi \rho_0 a^3 \left[ \ln \left( 1 + \frac{r}{a} \right) -
  \frac{r}{a(r+a)} \right]
\label{eq:nfw}
\end{equation}
where $a$ is the scale radius of the halo and $\rho_0$ is a characteristic
density. The scale radius $a$ depends on the concentration $c= r_{\rm h}/a$ of
the halo, where $5 \leq c \leq 15$ for halos of mass $10^{14} - 10^{10} \msun$
\citep[e.g.,][]{bullock.etal.2001}.

Elliptical galaxies and bulges that follow the $R^{1/4}$ law of
\cite{devaucoulers.1948} can usually be modelled by one of a family of
spherical density profiles characterised by an exponent $\gamma$
\citep{Dehnen93}. Here the enclosed mass goes as
\begin{equation}
M(r) = 4 \pi \rho_0 a^3 \left(\frac{r}{r+a}\right)^{3-\gamma}
\label{eq:gammas}
\end{equation}
with the Hernquist profile \citep{hernquist.1990} and the Jaffe cusp
\citep{Jaffe} corresponding to $\gamma = 1$ and $\gamma = 2$ respectively.

Referring now to Figure \ref{fig:r_sonic}, we can see that the sonic point is
in all these cases a significant fraction of the
halo radius at $T = T_{\rm vir}(r_{\rm h})$, and becomes larger than $r_{\rm
  h}$ for cooler gas. Thus, for a spherically-symmetric problem, the infalling
gas at or below the virial temperature (defined at $r_{\rm h}$)
\emph{must quickly tend to a free-fall solution} once inside the halo.

\bibliographystyle{mnras} 

\bibliography{bondi}

\end{document}